\documentclass[aps,prd,twocolumn,showpacs,preprintnumbers,nofootinbib,amsmath,amssymb,superscriptaddress]{revtex4}

\usepackage{placeins} 
\usepackage{float}
\usepackage{graphicx} 
\usepackage{dcolumn} 
\usepackage{bm} 
\usepackage{amsmath}
\usepackage[normalem]{ulem}

\newcommand{\beq}{\begin{eqnarray}}
\newcommand{\eeq}{\end{eqnarray}}
\newcommand{\tr}{\ensuremath{\mathrm{Tr}}}

\newcommand{\pbp}{\langle \bar \psi \psi \rangle}

\def\spose#1{\hbox to 0pt{#1\hss}}
\def\ltapprox{\mathrel{\spose{\lower 3pt\hbox{$\mathchar"218$}}
 \raise 2.0pt\hbox{$\mathchar"13C$}}}

\begin{document}

\title{
QCD phase diagram in a magnetic background for different values of the pion mass
}
\author{Massimo D'Elia}
\email{massimo.delia@unipi.it}
\affiliation{Universit\`a di Pisa, Largo B.~Pontecorvo 3, I-56127 Pisa, Italy}
\affiliation{INFN Sezione di Pisa, Largo B.~Pontecorvo 3, I-56127 Pisa, Italy}
\author{Floriano Manigrasso}
\email{floriano.manigrasso@pi.infn.it}
\affiliation{Universit\`a di Pisa, Largo B.~Pontecorvo 3, I-56127 Pisa, Italy}
\affiliation{INFN Sezione di Pisa, Largo B.~Pontecorvo 3, I-56127 Pisa, Italy}
\author{Francesco Negro}
\email{fnegro@pi.infn.it}
\affiliation{INFN Sezione di Pisa, Largo B.~Pontecorvo 3, I-56127 Pisa, Italy}
\author{Francesco Sanfilippo}
\email{sanfilippo@roma3.infn.it}
\affiliation{INFN Sezione di Roma Tre, Via della Vasca Navale 84, I-00146 Roma, Italy}

\date{\today} 

\begin{abstract}
We investigate the behavior of the pseudo-critical temperature
of $N_f = 2+1$ QCD as a function of a static magnetic background field
for different values of the pion mass, going up to $m_\pi \simeq 660$~MeV.
The study is performed {by lattice QCD simulations,} 
adopting a stout staggered discretization of 
the theory on lattices with $N_t = 6$ slices in the Euclidean temporal
direction; for each value of the pion mass the temperature
is changed moving along a line of constant physics.
We find that the decrease of $T_c$ as a function of $B$,
which is observed for physical quark masses, persists in the whole explored
mass range, even if the relative variation of $T_c$ appears to be a decreasing
function of $m_\pi$, approaching zero in the quenched limit. 
The location of $T_c$ is based on the renormalized 
quark condensate and its susceptibility{;}
 determinations based 
on the Polyakov loop lead to compatible results.
On the contrary,
inverse magnetic catalysis, i.e.~the decrease of the quark condensate as a function 
of $B$ in some temperature range around $T_c$, is not observed when the pion mass is high enough.
That supports the idea that inverse magnetic catalysis might be a secondary 
phenomenon, while the modifications induced by the magnetic background 
on the gauge field distribution and on the confining properties of the medium
could play a primary role in the whole range of pion masses.
\end{abstract}


\maketitle

\section{Introduction}
\label{intro}

The investigation of the properties of strong interactions
in a magnetic background field has been the subject of 
numerous studies in the recent past, see, e.g., 
Refs.~\cite{lecnotmag,Miransky:2015ava} for recent reviews.
Part of the interest is connected with phenomenology,
since strong background fields are expected in 
non-central heavy ion collisions~\cite{hi1, hi2, hi3, hi4, tuchin,
Holliday:2016lbx}, in some astrophysical objects like 
magnetars~\cite{magnetars}, and might have been produced 
during the cosmological electroweak phase
transition~\cite{vacha,grarub}. 
However, the issue is interesting also from a purely 
theoretical point of view, since background fields 
such that $e B \gtrsim \Lambda_{QCD}^2$ represent 
non-trivial probes of the non-perturbative properties
of strong interactions. Lattice QCD simulations
have been an essential tool to advance 
knowledge in this field, given the fact
that no technical issues, such as a sign problem, 
appear when one introduces a magnetic background
coupled to dynamical quark fields 
in the path-integral formulation of QCD.

{An} important aspect regards the influence of the magnetic field on the 
QCD phase diagram. 
Early lattice studies of $N_f = 2$ QCD, 
adopting unimproved
staggered fermions and {larger-than-physical} quark masses,
showed a slightly 
increasing behavior of the pseudo-critical 
temperature as a function of 
the magnetic field~\cite{demusa}.
 Those preliminary
indications however changed when new numerical simulations, 
exploiting an improved
discretization of $N_f = 2+1$ QCD with physical
quark masses, showed instead a substantial decrease of
$T_c$, of the order of 10-20\% for
 $|e| B \sim$ 1 GeV$^2$~\cite{reg0}. The reason for the discrepancy 
was ascribed to either the large quark masses, or possibly the large
cut-off effects present in the first study.

{
A decrease of the chiral pseudo-critical temperature with $T_c$
is counterintuitive, since it is expected 
on general grounds that the magnetic 
background enhances chiral symmetry breaking, 
a phenomenon known as magnetic catalysis}~\cite{salam,linde,Kawati:1983aq,Klevansky:1989vi,Suganuma,Klimenko,Schramm,
Klimenko:1993ec,Gusynin:1994re,Gusynin:1994xp,Shushpanov:1997sf,Shovkovy:2012zn}: 
{on that 
basis, the magnetic field should delay, rather than foster,
the restoration of chiral symmetry. Indeed, an increase
of the chiral restoration temperature was predicted by 
most models}~\cite{Fraga:2008qn,Boomsma:2009yk,Fukushima:2010fe,Gatto:2010pt,Kashiwa:2011js,Mizher:2010zb}, {with the exception
of the finite baryon density case~\cite{Preis:2012fh}.} 

Actually, while magnetic catalysis in the QCD vacuum (i.e.~at $T = 0$)
was confirmed by several lattice simulations~\cite{itep1,itep2,DEN,Ilgenfritz:2012fw,Bali:2012zg,Ilgenfritz:2013ara},
a new unexpected behavior was discovered for 
temperatures around $T_c$, consisting
in a decrease, rather than increase, 
of the quark condensate as a function of the magnetic 
field intensity~\cite{reg0,Bali:2012zg}.
{This phenomenon, later confirmed by further lattice 
studies}~\cite{Ilgenfritz:2013ara, Bruckmann:2013oba,Bornyakov:2013eya,Endrodi:2015oba},
was named as {\em inverse magnetic catalysis},
a name soon extended to indicate the decreasing behavior 
of $T_c (B)$ itself, thus assuming implicitly that
the latter is caused by the former.
{Many efforts have been done since then to interpret 
the new phenomenology within various 
model approaches}~\cite{icm0,icm1,Fraga:2012fs,Fukushima:2012xw,andersen1,andersen2,Fukushima:2012kc,Kojo:2012js,Endrodi:2013cs,
Chao:2013qpa,
icm2,Orlovsky:2013aya,Ferrer:2013noa,icm3,Kamikado:2013pya,Orlovsky:2013xxa,ferreira,Mueller:2014tea,Ruggieri:2014bqa,
Grunfeld:2014qfa,Farias:2014eca,Yu:2014sla,Ferreira:2014kpa,Ayala:2014iba,
Ayala:2014gwa,Tawfik:2014hwa,
dudal15,icm4,Cao:2014uva,Fayazbakhsh:2014mca,Yu:2014xoa,Braun:2014fua,Mamo:2015dea,
Mueller:2015fka,Rougemont:2015oea,Costa:2015bza,
icm5,mao,Tawfik:2016lih,Farias:2016gmy,Evans:2016jzo,Li:2016gtz,
Tawfik:2016gye, Fang:2016cnt, Fukushima:2016vix,Ruggieri:2016xww,
Pagura:2016pwr,Li:2016gfn,Gursoy:2016ofp,Dudal:2016joz,Fu:2017vvg,
Gursoy:2017wzz,GomezDumm:2017iex,Rodrigues:2017iqi,Rodrigues:2018pep}.

{Whether the chiral properties of QCD are at the 
origin of the decreasing behavior of $T_c(B)$ or not is still unclear.
The behavior has been also associated 
(see, e.g., Refs.}~\cite{icm0,Allen:2013lda,Ballon-Bayona:2013cta,Anber:2013tra, Ballon-Bayona:2017dvv})
{with the paramagnetic properties of the Quark-Gluon Plasma (QGP) 
phase, which have clearly emerged from several lattice QCD 
studies}~\cite{Bonati:2013lca,Levkova:2013qda,Bonati:2013vba,Bali:2013owa,Bali:2014kia}. 
Moreover, the magnetic field affects
many properties of the gluon 
fields~\cite{Bruckmann:2013oba,Ilgenfritz:2013ara,Galilo:2011nh,DElia:2012ifm,ozaki,Bali:2013esa,Ayala:2015qwa,DElia:2015eey} 
through the indirect coupling with them
induced by dynamical quarks. 
Many effects take place
directly at the level of the confining 
properties of the theory~\cite{anisotropic,Bonati:2014ksa,Rougemont:2014efa,Simonov:2015yka,Bonati:2016kxj,Bonati:2017uvz,Bonati:2018uwh},
leading to anisotropies in the static quark-antiquark
potential at $T = 0$
and to a suppression
of the string tension close to $T_c$~\cite{Bonati:2016kxj}.

For this reason, one may ask whether 
the decrease of $T_c$ could not be ascribed to 
these effects and thus be a sort of {\em deconfinement catalysis},
with the decrease of the quark condensate being just a secondary
effect, induced by the fact that the magnetic background
fosters deconfinement, hence chiral symmetry restoration.
Actually, in view of the strict entaglement between
confinement and chiral symmetry breaking, which 
is not yet fully understood, the question of understanding
which is the true driving phenomenon could be ill-posed.

However, from a practical point of view, we can ask how the situation
changes as the quark mass spectrum is changed. In particular,
adopting substantially larger than physical
quark masses, i.e.~approaching the quenched limit, 
one can explore a regime where
chiral symmetry ceases to be a {good} 
symmetry of the theory,
while symmetries associated to confinement, like center symmetry,
become more and more {relevant}. Then, one can try
to asnwer some some clear-cut questions:
\begin{itemize}
\item[{\em i)}] is the decrease of $T_c$ as a function of $B$ still
observed?
\item[{\em ii)}] in case it is, is the decrease always associated with 
inverse magnetic catalysis?
\end{itemize}
\noindent
In this study we try to answer these questions adopting 
the same discretization of $N_f = 2+1$ used in Ref.~\cite{reg0}
and exploring three different lines of constant physics, corresponding
to pseudo-Goldstone pion masses $m_\pi = 343, 440$ and 664 MeV on lattices 
with $N_t = 6$. 
Anticipating our main results, the answer to the first 
question is yes, while that to the second question is no.

This is of course also important to understand the reason of the 
discrepancy observed in early lattice studies where
a slight increase of $T_c$, instead of a decrease, was observed.
The unphysical quark mass spectrum adopted in Ref.~\cite{demusa}
was considered as one possible reason~\cite{reg0}.
Our present results suggest that the different behavior 
can be likely ascribed to large lattice artifacts induced
by the coarse lattice spacing and unimproved discretization
adopted in Ref.~\cite{demusa}.
That is in agreement {with} 
recent studies~\cite{Tomiya:2017cey}, 
adopting the same unimproved discretization
of Ref.~\cite{demusa}, where the increase of $T_c$ as a function
of $B$ is observed even for smaller than physical quark masses.

The paper is organized as follows. In Sec.~\ref{setup} we describe
our lattice discretization of $N_f = 2+1$ QCD with a 
magnetic background, as well as the physical observables
and the numerical setup to work on lines of constant physics.
In Sec.~\ref{tc_b0} we present results regarding the behavior 
of the pseudo-critical temperature as a function of 
the pion mass at $B = 0$, while Sec.~\ref{tc_nonzerob}
contains our main results obtained in the presence 
of a magnetic background. Finally, in Sec.~\ref{catalysis}
we investigate the fate of inverse magnetic catalysis 
in the large quark mass limit and
in Sec.~\ref{conclusions} we draw our conclusions.

\section{Numerical Methods}
\label{setup}

In this {study} 
we investigated $N_f=2+1$ QCD at different values
of the quarks masses, adopting a setup with two degenerate
up and down flavors,
$m_u=m_d=m_\ell$,
and a strange-to-light mass ratio fixed at {its physical 
value} $m_s/m_\ell=28.15$.
We adopted the stout-smearing improved staggered quark action
and the tree-level Symanzik improved gauge 
action~\cite{Weisz:1982zw, Curci:1983an}.
With this choice the {discretized} partition function takes the form
\begin{equation}\label{eq:partfunc}
Z(m_\ell,B) = \int \!\mathcal{D}U \,e^{-S_{Y\!M}}
\!\!\!\!\prod_{f=u,\,d,\,s} \!\!\!
\det{({M_{\textnormal{st}}^f[m_f,B]})^{1/4}}\ ,
\end{equation}
where $\mathcal{D}U$ is the functional integration over all the
possible $SU(3)$ gauge field configurations.
The gauge action $S_{Y\!M}$ is written in terms of the real part
of the trace of the $1\times 1$ and $1\times 2$ Wilson loops
(respectively denoted as $P_{i;\mu\nu}^{1\times 1}$ and $P_{i;\mu\nu}^{1\times 2}$):
\begin{equation}\label{eq:tlsyact}
S_{Y\!M}= - \frac{\beta}{3}\sum_{i, \mu \neq \nu} \left(
\frac{5}{6} P^{1\!\times \! 1}_{i;\,\mu\nu} - \frac{1}{12}
P^{1\!\times \! 2}_{i;\,\mu\nu} \right)\ .
\end{equation}
The Dirac operator $M_{\textnormal{st}}^f[m_f,B]$ related
to the flavor $f$ which appears in $Z(m_\ell,B)$ takes the form
\begin{equation}\label{eq:fermionmatrix}
\begin{aligned}
(M^f_{\textnormal{st}})_{i,\,j} =\ & am_f
  \delta_{i,\,j}+\!\!\sum_{\nu=1}^{4}\frac{\eta_{i;\,\nu}}{2}
  \left(u^f_{i;\,\nu}U^{(2)}_{i;\,\nu}\delta_{i,j-\hat{\nu}}
  \right. \nonumber\\ &-\left. u^{f*}_{i-\hat\nu;\,\nu}U^{(2)\dagger}_{i-\hat\nu;\,\nu}\delta_{i,j+\hat\nu}
  \right)\ ,
\end{aligned}
\end{equation}
where the staggered phases are denoted by $\eta_{i;\,\nu}$
and where $U^{(2)}_{i;\,\mu}$ is the two times
stout-smeared~\cite{Morningstar:2003gk} $SU(3)$ link variable
(with an isotropic smearing parameter $\rho=0.15$) at position
$i$ pointing along the direction $\mu$.
The configuration of $U(1)$ link variables $u^f_{i;\,\mu}$ is
fixed at the beginning of each simulation in order to reproduce
the desired value of {the} external uniform magnetic field $B_z$.
{The only nontrivial phases are:}
\begin{eqnarray}\label{eq:u1field}
u^f_{i;\,y}=e^{i a^2 q_f B_{z} i_x} \ , \quad
{u^f_{i;\,x}|}_{i_x=N_x}=e^{-ia^2 q_f N_x B_z i_y}\, ,
\end{eqnarray}
where the charge of each flavor $f$ is denoted as $q_f$ ($q_u=-2q_d=-2q_s=2e/3$).
This expression has been derived
to describe the four-potential of a uniform magnetic field
over a manifold with periodic boundary conditions,
{such} as the lattice discretized space we adopt.
Such periodicity constrains $B_z$ to quantized 
values~\cite{tHooft:1979rtg, Damgaard:1988hh, AlHashimi:2008hr}
\begin{equation}\label{eq:bfield}
eB_z={6 \pi b}/{(a^2 N_x N_y)}\ ,
\end{equation}
where $b$ is integer valued.
The partition function is periodic in $b$ with
{period $N_x N_y$.}

{Numerical simulations have been performed using 
the Rational Hybrid Monte-Carlo algorithm (RHMC)}~\cite{rhmc1}
{implemented in the NISSA code}~\cite{nissa}
{and in the OpenStaPLE code for GPUs}~\cite{gpu2,gpu3}. 
{We have performed around 100 runs with different combinations
of $T$ and $B$ for each value of the pion mass, with average statistics
of approximately 3000 RHMC trajectories for each run.}

\subsection{Lines of constant physics}

To study the chiral symmetry restoration crossover
at different values of the pion mass $m_\pi$, we need to perform,
for each chosen value of $m_\pi$, several simulations
at different {values of the 
temperature $T = 1/(N_t a)$, i.e.~at different lattice spacings $a$,
since we work at fixed $N_t = 6$.} 
This requires the preliminary knowledge of the lines in the
$\beta-m_\ell$ plane along which the pion mass stays constant
{at the chosen values ($m_\pi = 343, 440$ and $664$ MeV),} which we
refer to as {\it lines of constant physics} (LCP).

To achieve this, we performed {preliminary} 
$T=0$ numerical simulations on
a $24^3\times 32$ lattice at {7} values of $\beta$
($\beta = 3.45, 3.55, 3.62, 3.73, 3.85, 4.00, 4.10$)
and at 5 - 7 values of $m_\ell$ for each $\beta$
(in the range $0.007 \lesssim m_\ell \lesssim 0.19$),
for a total of $42$ simulation points.
Then, we associated each simulation point with a value
of the lattice spacing $a$ by adopting the $w_0$ scale
setting approach~\cite{wflow_2}, 
based on the gradient flow technique~\cite{wflow_1},
{and assuming that $w_0$ is independent of $m_\pi$.}
Moreover, we extracted the lightest pseudoscalar
pion mass $m_\pi$ from the decay of Euclidean time
correlators of the appropriate staggered quark operators.
The obtained values of $a$ and $m_\pi$ {extend} from
$\sim\!0.07$~fm to $\sim\!0.3$~fm and from $\sim\!300$~MeV
to $\sim\!700$~MeV respectively, covering completely
the parameter region we are interested in.

\begin{figure}[t!]
\begin{center}
\includegraphics*[width=1\columnwidth]{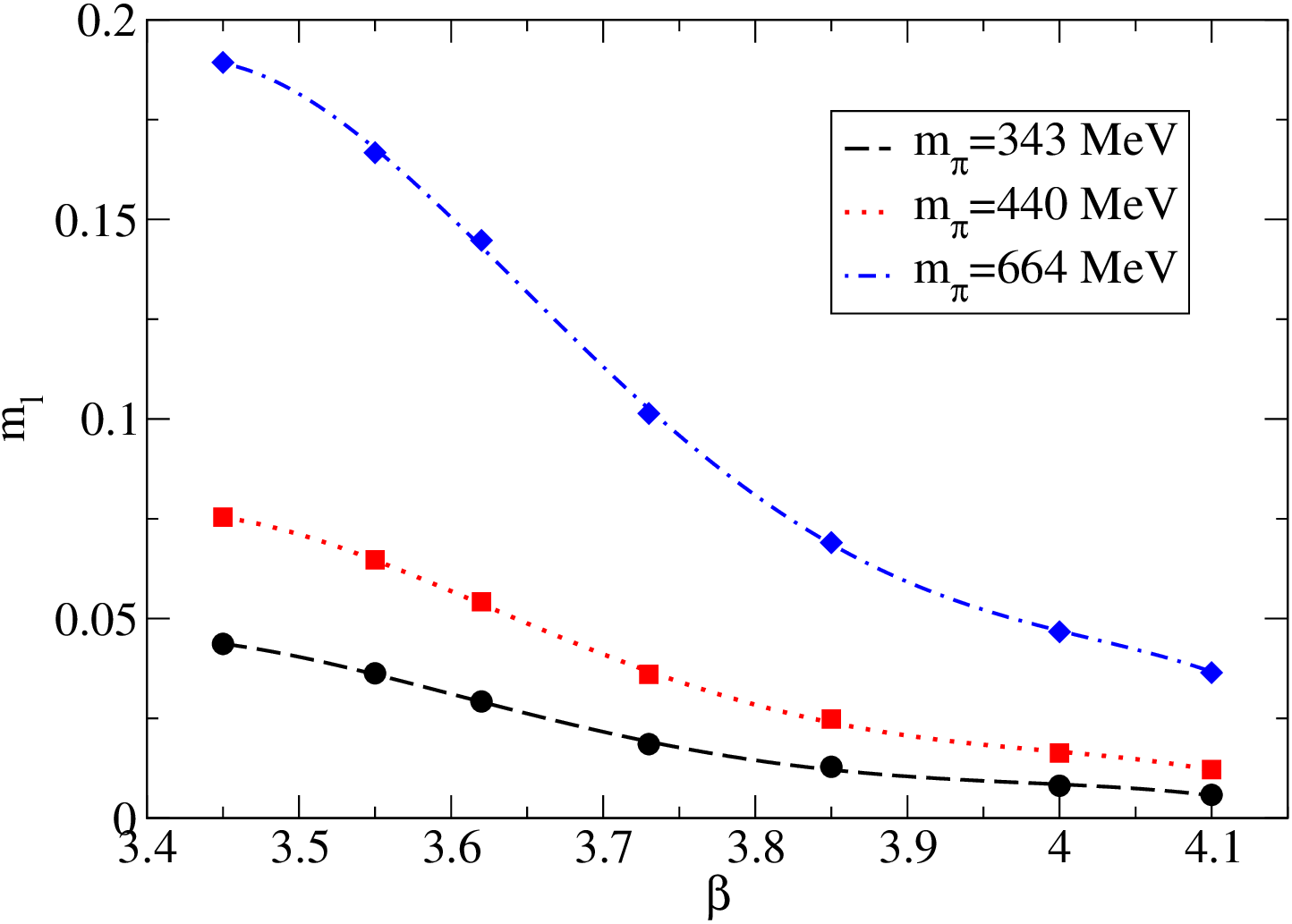}
\end{center}
\caption{Lines of constant pion mass in the $m_\ell-\beta$ plane.}
\label{fig:lcp1}
\end{figure}

At fixed $\beta$, we interpolated the pion mass $m_\pi$
as function of $m_\ell$ with a 3rd order polynomial.
This allows us to determine, {for each $\beta$},
 the bare quark mass $m_\ell$
that corresponds to the chosen {value of the pion mass}.
The LCP 
are obtained
by interpolating $m_\ell$ as a function of $\beta$ with a 4th order polynomial,
as shown in Fig.~\ref{fig:lcp1}.

\begin{figure}[t!]
\begin{center}
\includegraphics*[width=1\columnwidth]{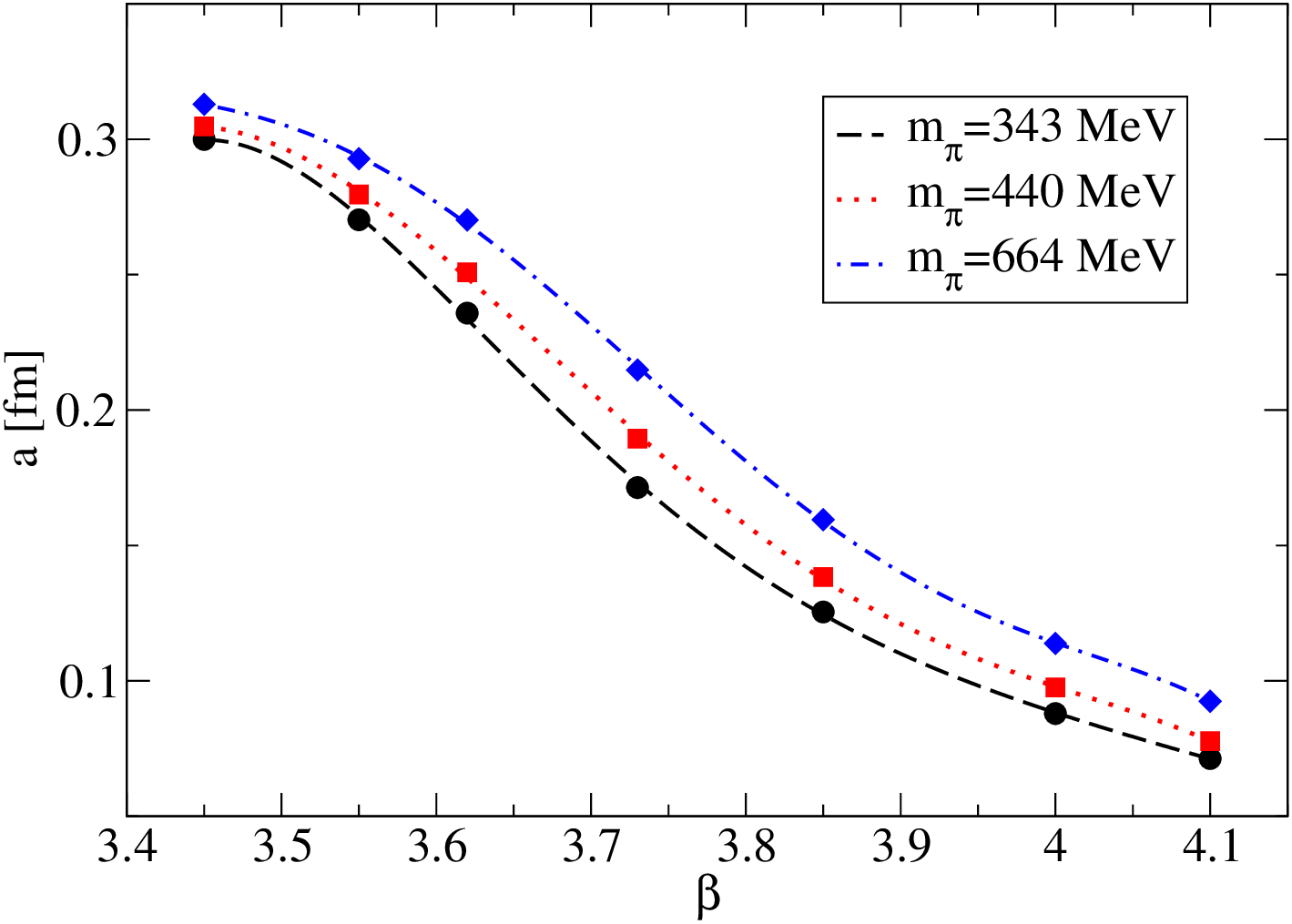}
\end{center}
\caption{Dependence of the lattice spacing $a$ on $\beta$ along the lines of constant pion mass.}
\label{fig:lcp2}
\end{figure}

For the determination of the lattice spacing along
these lines we adopted a similar procedure.
Again, at fixed {$\beta$}, we interpolated the
dependence of the lattice spacing $a$ on the
pion mass $m_\pi$ {with} the function
\begin{equation}\label{eq:fitexpmpi}
a(m_\pi) = A(1+B\, e^{-C\, m_\pi}),
\end{equation}
which approaches a finite value as we go
towards the quenched limit $m_\pi\to\infty$.
{In this way we extract the value of $a$ at 
the chosen value of the pion mass
for all the explored $\beta$ values.}
Then we give an estimate of the lattice spacing along
the LCP by interpolating
these data {by a 5th order polynomial\footnote{Actually, both 
for $a(\beta)$ and $m_l (\beta)$ a 4th order or a 5th order polynomial
work equally well, yielding values which differ,
in the interpolated regions,
by far less than 1\%.}
in $\beta$}.
The results of these interpolations are shown in Fig.~\ref{fig:lcp2}.

\subsection{Physical observables}

For the purpose of this study and, in particular,
for the discussion of the fate of (inverse) magnetic catalysis,
we focused 
on the {quark} condensate and its susceptibility.
The {quark} condensate of the flavor $f$ is expressed as
\begin{eqnarray}\label{eq:fcond}
&\Sigma&\!\!\!\!_f(T,B) = \frac{T}{V}\frac{\partial \log Z}{\partial m_f} = \frac{T}{VZ(B)}  \int \!\mathcal{D}U\,e^{-S_{Y\!M}}\
\cdot
\\\nonumber &\phantom{a}&\!\!\!\!\!\!\!\!\!\!\!\!\!\!\!\!\!\!\cdot{\rm Tr}((M^f_{st}[m_f,B])^{-1})\,
\!\!\!\!\prod_{f'=u,\,d,\,s} \!\!\!
\det{({M_{\textnormal{st}}^{f'}[m_{f'},B]})^{1/4}}
\end{eqnarray}
where $T$ is the temperature, 
$V$ is the spatial volume 
{and the trace is evaluated, as usual, by means of noisy
estimators.}
At non-zero magnetic field, the two light quarks condensates
differ because of their electric charge: they couple in
a different way to the magnetic field.
We introduce the light quark condensate $\Sigma_\ell$ defined as
\begin{equation}\label{eq:lcond}
\Sigma_\ell(T,B) = \Sigma_u(T,B) + \Sigma_d(T,B).
\end{equation}
This observable is affected by both additive and
multiplicative renormalizations. As it has been pointed
out in Ref.~\cite{reg0}, the presence of the external magnetic
field does not introduce new $B-$dependent divergencies.
For this reason, the renormalization prescription introduced
in Ref.~\cite{Endrodi2011}, which exploits $T = 0$ quantities
to perform the additive renormalization and the value of
the bare quark mass to take care of multiplicative ones,
can be extended to the $B \neq 0$ case as
\begin{equation}\label{eq:ren_pres_wupp}
\Sigma_\ell^r (T,B) = \frac{m_\ell}{M_\pi^4}\left(\Sigma_\ell(T,B)-\Sigma_\ell(0,0)\right),
\end{equation}
where the two condensates are computed at the same UV cutoff
{(i.e.~the same bare parameters).}

The location of $T_c$ is usually defined, in terms of the renormalized
light condensate, as the point of maximum slope, i.e.~the point
where $\Sigma_\ell^r$ has an inflection point
as a function of $T$ and
the absolute value of
$\partial \Sigma_\ell^r / \partial T$ reaches a maximum.
Alternatively, one can consider the
behavior of the chiral susceptibility, i.e.
\begin{equation}\label{eq:chirsusc}
\chi_{\ell}= \frac{\partial \Sigma_\ell}{\partial m_\ell} .
\end{equation}
For $\chi_\ell$ the renormalization is perfomed
{in a similar way,}
\begin{equation}\label{eq:chirsuscren}
\chi_{\ell}^r(T,B)= m_\ell^2\left( \chi_\ell(T,B)-\chi_\ell(0,0) \right) \, ,
\end{equation}
{and we look for the peak of the dimensionless
ratio $\chi_\ell^r(T,B)/m_\pi^4$ to locate $T_c$.}
For the renormalization of both {$\Sigma_\ell$ and $\chi_{\ell}$}
we have exploited the same zero temperature
runs that we used to determine the LCP. 

In Ref.~\cite{DEN} it was observed that the
change in the quark condensate due to the magnetic
field can be ascribed both to {\it sea}
(dynamical in Ref.~\cite{DEN}) and to {\it valence} quarks effects.
At $T=0$, {and for small enough magnetic fields,} 
the two contributions add approximately
to the total change of the condensate,
{with a {sea}
contribution amounting
to $\sim 30\%$ of the total signal.}
{The sea contribution} is particularly interesting because it is only
related to {the modification of the gauge configurations
which are mostly relevant in the path integral.} 
Moreover, it was observed~\cite{Bruckmann:2013oba} that {around} $T_c$, 
the sea contribution {changes sign},
possibly resulting in {the observed} overall decrease of
the condensate {known as}
inverse magnetic catalysis.
It is then {useful} to introduce the sea quark condensate,
defined as the standard condensate but with
the external field switched off in the observable:
\begin{eqnarray}\label{eq:seqcond}
&\Sigma&\!\!\!\!_f^{sea}(T,B) = \frac{T}{VZ(B)}  \int \!\mathcal{D}U\,e^{-S_{Y\!M}}\cdot
\\\nonumber &\phantom{a}&\!\!\!\!\!\!\!\!\!\!\!\!\!\!\!\!\!\!\cdot{\rm Tr}((M^f_{st}[m_f,0])^{-1})\,
\!\!\!\!\prod_{f'=u,\,d,\,s} \!\!\!
\det{({M_{\textnormal{st}}^{f'}[m_{f'},B]})^{1/4}}.
\end{eqnarray}
The sea contribution to the change of the light
condensate at finite $B$ is then expressed as
\begin{equation}\label{eq:deltaseacond}
\Delta\Sigma^{sea}_\ell(T,B) 
= \Sigma^{sea}_u(T,B)+\Sigma^{sea}_d(T,B)-\Sigma_\ell(T,0) \, .
\end{equation}

{Finally, as a further observable
to locate the crossover,} 
we introduce the unrenormalized Polyakov loop 
\begin{equation}\label{eq:polyloop}
P(T,B) = \frac{1}{V} \langle \sum_{i_x,i_y,i_z}\textrm{Re}\tr\prod_{i_t=0}^{N_t-1} U_{i,t} \rangle \, ,
\end{equation}
{which is more directly connected to the 
confining properties of strong interactions
and becomes an exact order parameter
in the infinite mass limit that we are somehow approaching. 
As for the quark condensate, we will look for its inflection
point to locate $T_c$.}

\subsection{Determining observables 
as a function of $T$ at fixed values of the magnetic field}

{When moving along a LCP, one would like to keep the magnetic field
strength constant as well. Since $e B$ is given by
Eq.~(\ref{eq:bfield}), one should tune the parameter $b$
to balance the change of the lattice spacing along the LCP;
however, being $b$ integer-valued, this is not 
possible in practice.}
We have approached this problem similarly to Ref.~\cite{reg0}: 
{simulations have been performed at various temperatures
and, for each $T$, at various values of $b$.
Then, to reach the desired value of $eB$, observables 
have been interpolated by a cubic spline in $b$,
arranged in such a way that its
first derivative is zero at $b=0$, as expected by analyticity 
in $eB$ and charge conjugation symmetry}.

\begin{figure}[t!]
\begin{center}
\includegraphics*[width=1\columnwidth]{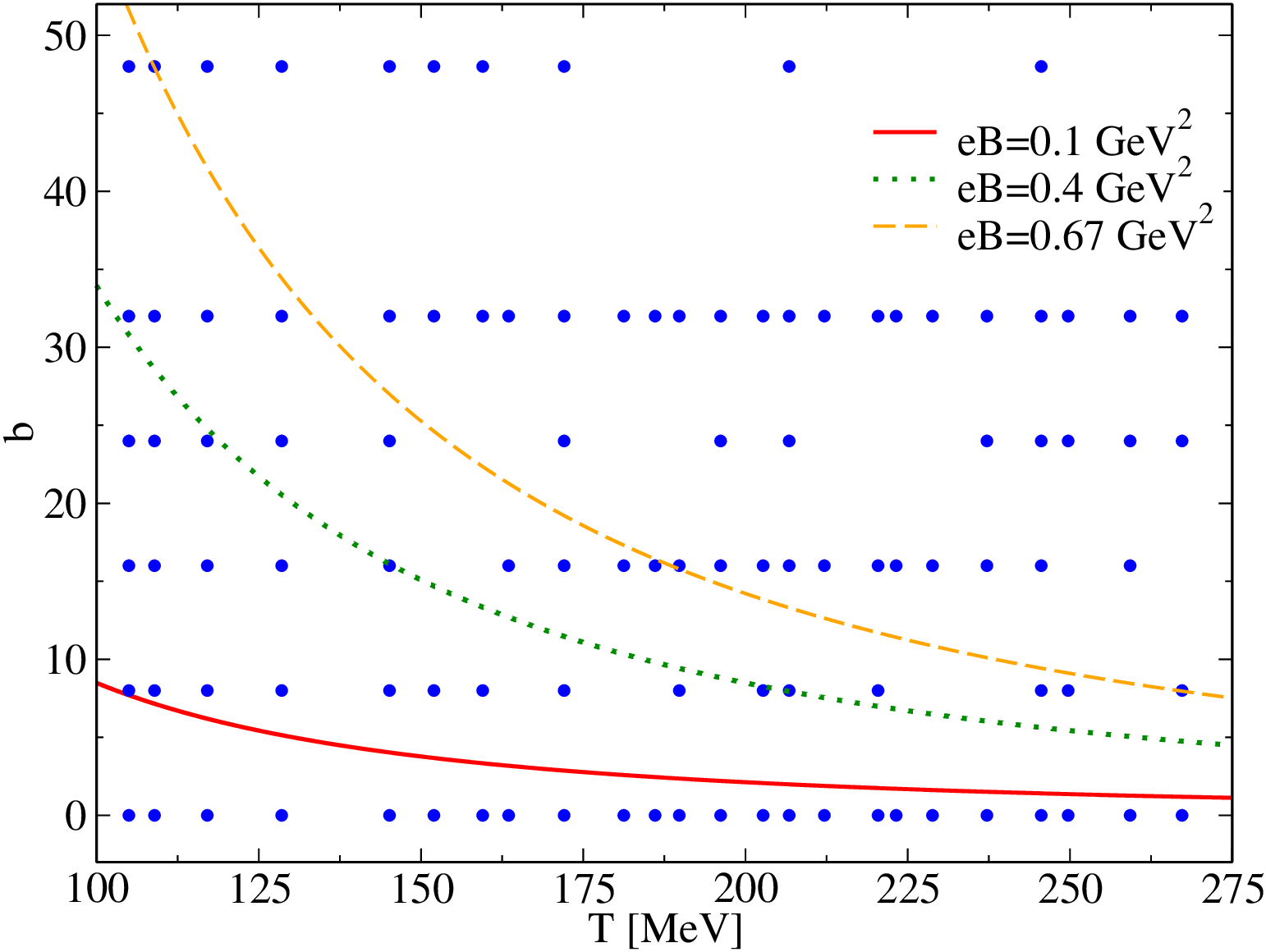}
\end{center}
\caption{Lines of constant $B$ together with simulation points (blue dots) on $24\times 6$ lattices at $m_\pi=664\,\text{MeV}$.
}
\label{fig:1}
\end{figure}

\begin{figure}[t!]
\begin{center}
\includegraphics*[width=0.95\columnwidth]{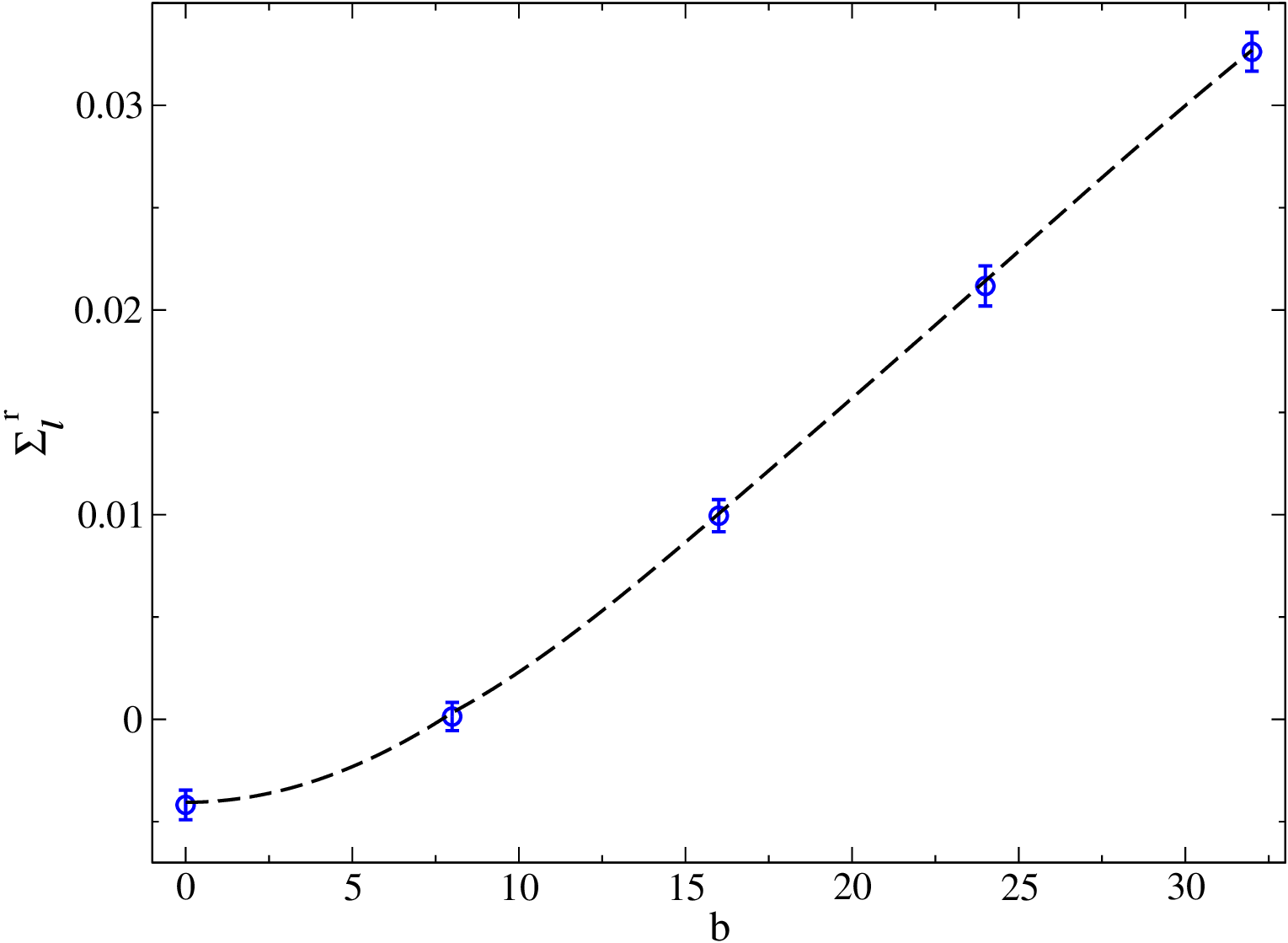}
\end{center}
\caption{{Light quark} condensate at $T=141\,\text{MeV}$ for $m_\pi=343\,\text{MeV}$, together with a cubic spline.
}
\label{fig:2}
\end{figure}

We report in Fig.~\ref{fig:1} the explored grid of such 
simulation points for $m_\pi=664$ MeV, together with the lines 
{of constant $eB$ (in particular, $eB=0.1,\ 0.4,\ 0.67\ {\rm GeV}^2$).}
Such lines are characterized by the equation
\begin{equation}\label{eq:lcb}
b = \frac{eB}{T^2}\cdot\frac{ N_xN_y}{6\pi N_t^2} \, .
\end{equation}
A typical interpolation of the {quark} condensate $\Sigma_\ell$ is shown in Fig.~\ref{fig:2}, where data points corresponds to $\
m_\pi=343$ MeV at $T=141$ MeV.

\section{Pseudo-critical temperature at $B = 0$}
\label{tc_b0}

\begin{figure}[t!]
\centering
\includegraphics*[width=1\columnwidth,height=1\columnwidth]{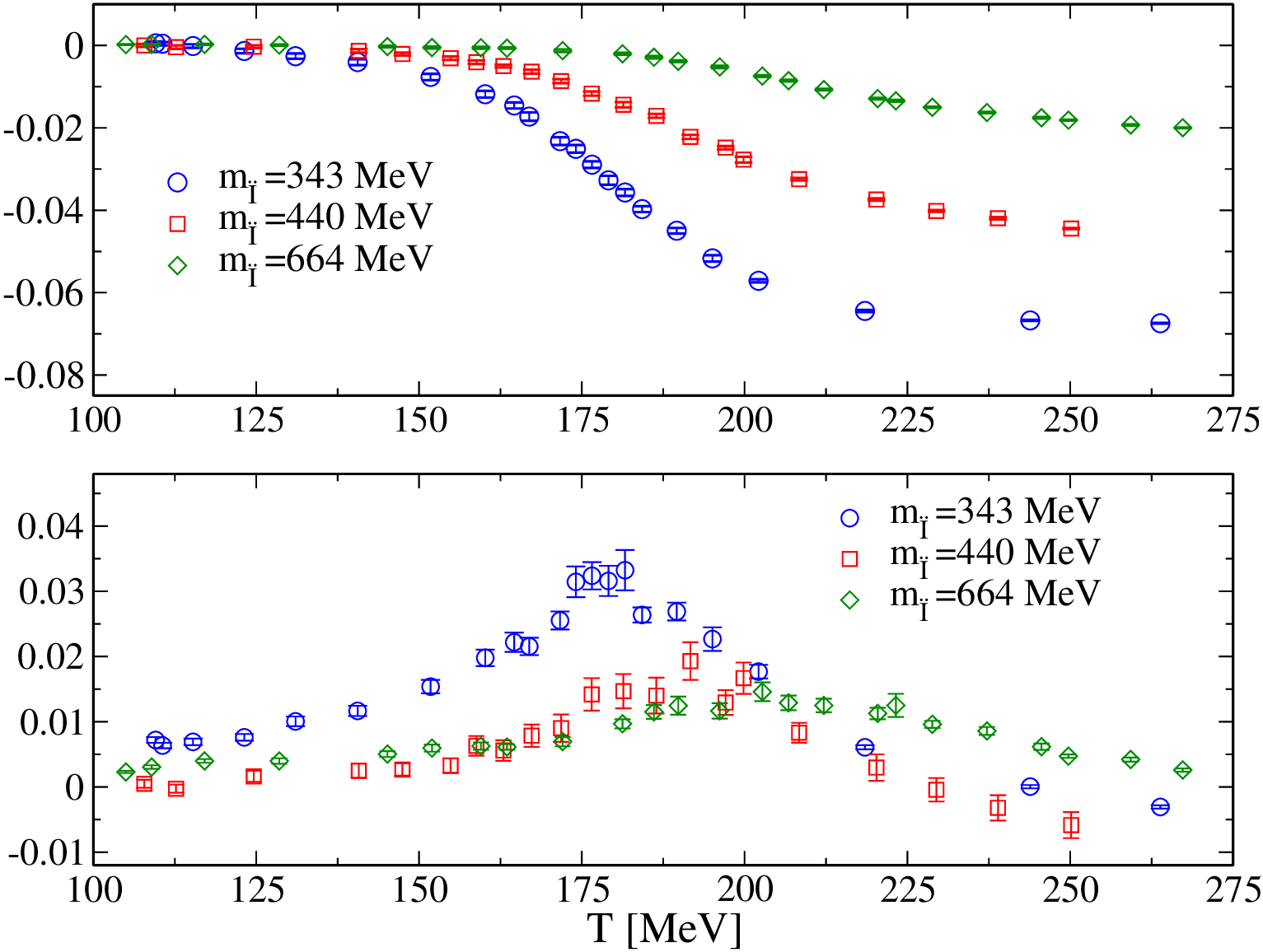}
\caption{{Renormalized light quark} 
condensate and susceptibility as a function of $T$ at $B=0$ for different
values of the pion mass.
}
\label{fig:3}
\end{figure}

We illustrate our results starting from the determination
of the pseudo-critical temperature at $B = 0$,
which represents the starting input
to study the dependence of $T_c$ on $B$.
In Fig.~\ref{fig:3}, the renormalized {quark} condensate 
and its susceptibility are shown as a function of 
$T$ for the different explored values of $m_\pi$.
The peak of the chiral susceptibility, as well as
the inflection point of the {quark} condensate,
clearly shift to higher temperatures as $m_\pi$
increases; at the same time, 
a general weakening of the crossover strength is observed, which is  
consistent with the fact that
chiral symmetry restoration is less
and less relevant to strong interaction dynamics as
the quenched limit is approached.

In order to determine $T_c$, the inflection point
of the renormalized condensate has been located
by fitting data
according to an arctangent or a cubic polynomial
ansatz, while
the maximum of the susceptibility peak has been obtained
by fitting data according to a quadratic or a Lorentzian 
function of $T$. In both cases,
systematic errors have been estimated
by varying the range of fitted data points or the fitting function.
 Results obtained from the inflection point of the condensate are reported in
Table~\ref{tab:tc_mpi}.

\begin{table}
	\caption{\label{tab:tc_mpi} Pseudo-critical temperature at $B=0$ at different pion masses.}
	\begin{ruledtabular}
		\begin{tabular}{ccc}
		$m_{\pi}\,[\text{MeV}]$& $T_C\,[\text{MeV}]$ & $\Delta T_C\,[\text{MeV}]$ \\ 
						\hline
		342 & 180.5 & 2  \\
		440 & 191 & 2	\\		
		664 & 209 & 2 
		\end{tabular}
	\end{ruledtabular}
\end{table}

\begin{figure}[b!]
\centering
\includegraphics*[width=1.0\columnwidth]{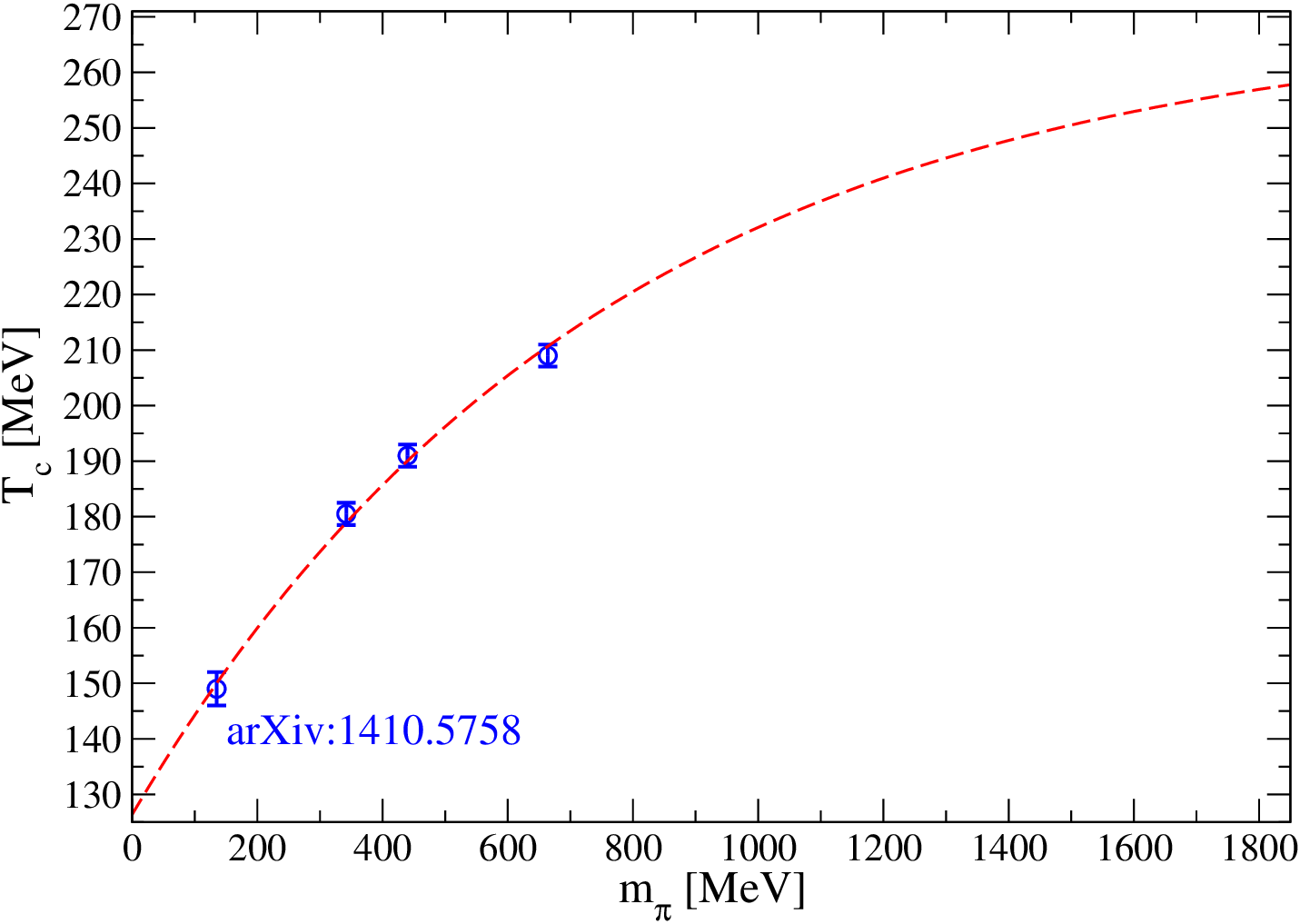}
\caption{Pseudo-critical temperature at $B=0$ as a function of $m_\pi$.
The dashed line represents the result of a best fit 
according to Eq.~(\ref{chiral_quenched_limit_Tc}).
}
\label{fig:4}
\end{figure}

It is interesting to consider how $T_c$ depends on $m_\pi$. To that 
aim, in Fig.~\ref{fig:4} we plot the results reported 
in Table~\ref{tab:tc_mpi} together with a determination
at the physical point adopting the same discretization
on $N_t = 6$ lattices, $T_c(m_\pi = 135\, {\rm MeV}) = 149(3)$, 
that we have taken
from Ref.~\cite{Bonati:2014rfa}.
On general grounds, one expects that $T_c$ approaches finite values
both in the quenched limit, $\lim_{m_\pi \to \infty} T_c \equiv T_c^{quench}$,
and in the chiral limit, $\lim_{m_\pi \to 0} T_c \equiv T_c^{\chi}$.
Moreover, approaching the chiral limit, one
expects 
\beq
T_c(m_\pi) = T_c^{\chi} + A\, m_\pi^{2/(\beta\delta)}
\label{tcchirlim}
\eeq
where $\beta$ and $\delta$ are the critical indexes describing
the critical behavior 
around the chiral point, even if that could be
not relevant to our range of pion masses,
which is far away from the chiral limit.
A previous study of the dependence 
of $T_c$ on $m_\pi$ has been reported in Ref.~\cite{Karsch:2000kv},
adopting a different staggered discretization and 
a range of large pion masses similar to ours: in 
that case it was found that 
$T_c(m_\pi)$ behaves more or less linearly in $m_\pi$
when approaching the chiral limit,
a result which is not far from the prediction
of Eq.~(\ref{tcchirlim}) for various universality
classes which could be relevant 
in the case of a second order chiral transition (e.g., $O(2)$ or $O(4)$), 
$2/(\beta\delta) \sim 1.2$.

Inspired by these considerations and previous findings, 
we have tried to fit the data reported in 
Fig.~\ref{fig:4} according to
\beq
T_c(m_\pi) = T_c^{quench} - (T_c^{quench} - T_c^{\chi})\, \exp(-m_\pi/M)
\label{chiral_quenched_limit_Tc}
\eeq
and, fixing $T_c^{quench} = 270$ MeV, we obtain 
$T_c^{\chi} = 128(4)$ MeV and 
$M = 763(39)$ MeV with $\tilde \chi^2 = 1.54/2$. 
The value obtained for the critical temperature
in the chiral limit, $T_c^\chi \equiv T_c (m_\pi = 0)$, is not
unreasonable, given the preliminary estimate 
$T_c^{\chi} = 138(5)$ MeV in the continuum limit recently 
reported in Ref.~\cite{Ding:2018auz}, and given that our 
estimate represents a
very rough extrapolation from a region of large pion masses.

The functional dependence in Eq.~(\ref{chiral_quenched_limit_Tc})
is by no means unique: various other functions, sharing 
the same rough properties exposed above, fit data equally well, like
for instance 
$$T_c(m_\pi) = T_c^{\chi} + \frac{2}{\pi}(T_c^{quench} - T_c^{\chi})\, \arctan(m_\pi/M)$$ or
$$T_c(m_\pi) = T_c^{\chi} + (T_c^{quench} - T_c^{\chi})\, 
\frac{m_\pi^B}{A + m_\pi^B}$$ 
with $B \sim 1$\footnote{Taking
$m_\pi^B$ with $B \neq 1$ as an argument, which could accomplish 
for the possible critical behavior around the chiral point,
works well also for the $\arctan$ and the exponential ansatz.}.

\section{Pseudo-critical temperature at non-zero magnetic field}
\label{tc_nonzerob}

\begin{figure}[t!]
\centering
\includegraphics*[width=1\columnwidth]{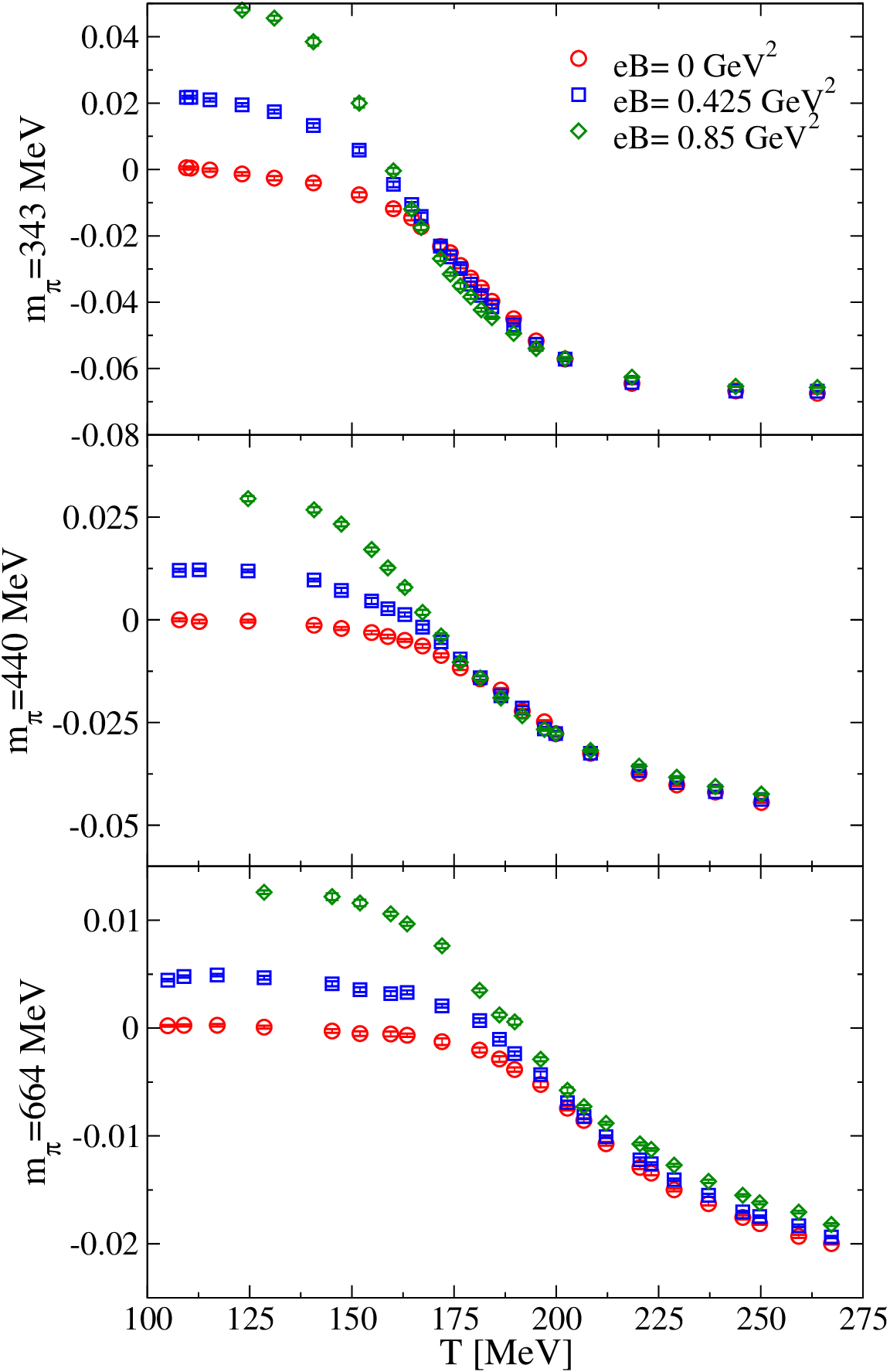}
\caption{Renormalized condensate for different values of the 
magnetic field at the different pion masses.
}
\label{fig:cond_T_B}
\end{figure}

\begin{figure}[h!]
\centering
\includegraphics*[width=1\columnwidth]{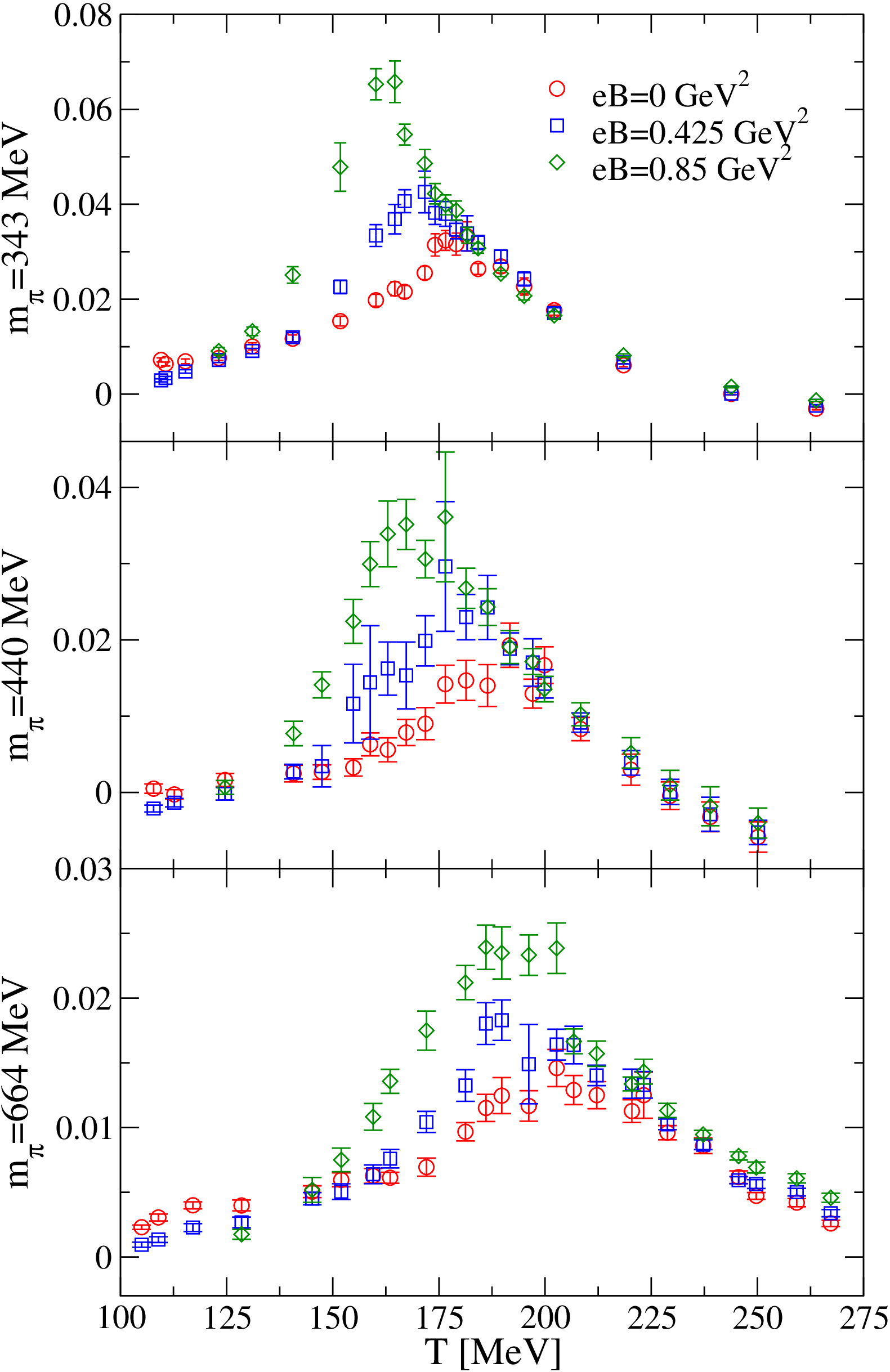}
\caption{Renormalized chiral susceptibility for different values of the 
magnetic field at the different pion masses.
}
\label{fig:susc_T_B}
\end{figure}

\begin{figure}[t!]
\centering
\includegraphics*[width=1\columnwidth]{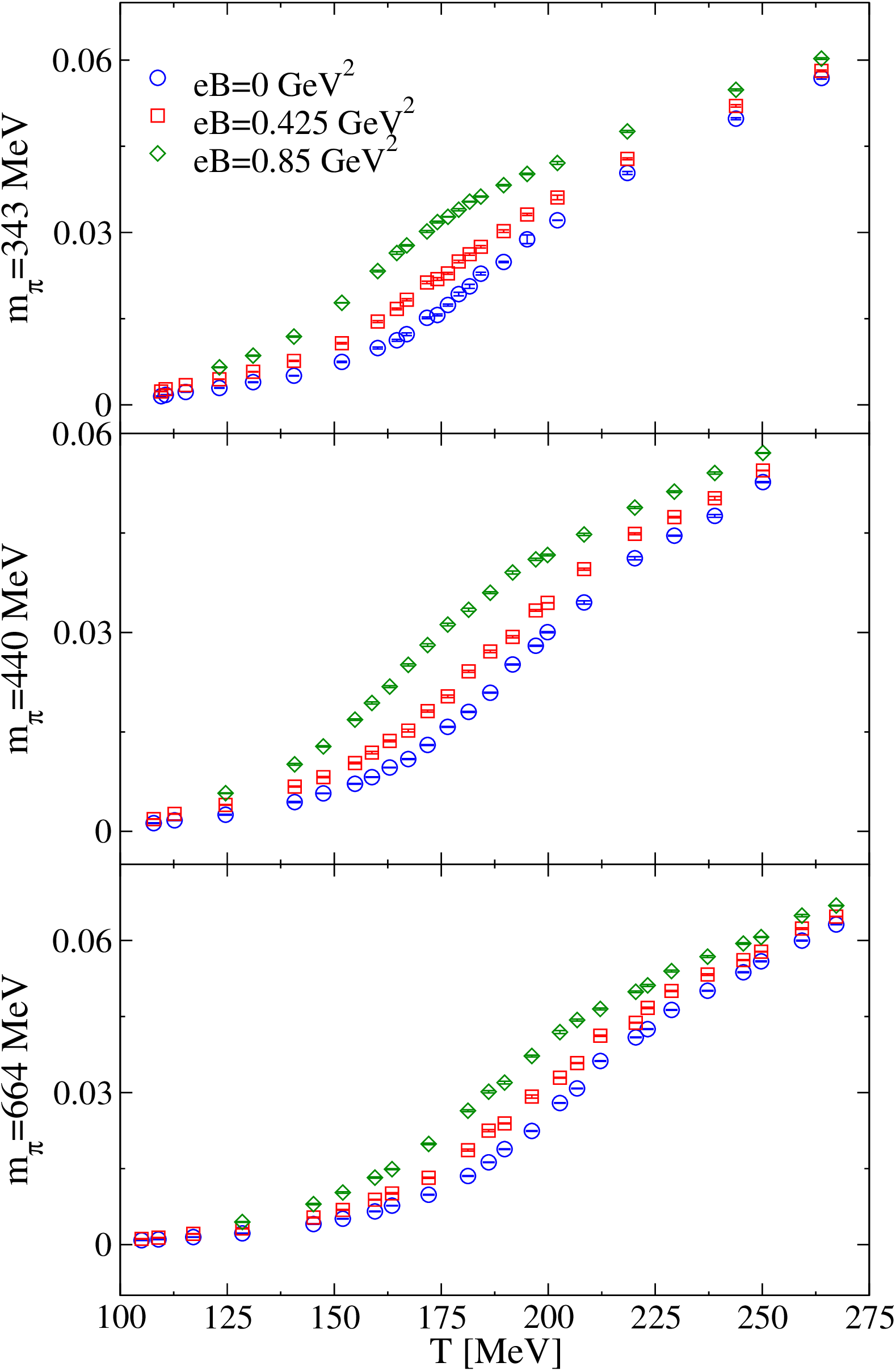}
\caption{Unrenormalized Polyakov loop for different values of the 
magnetic field at the different pion masses.
}
\label{fig:poly_T_B}
\end{figure}

The main result of our study can be already 
appreciated by looking at 
Figs.~\ref{fig:cond_T_B} and \ref{fig:susc_T_B}, where 
we plot respectively the renormalized condensate 
and the renormalized chiral susceptibility as a function
of the temperature for different values of the magnetic 
background and of the pion masses. The inflection
point of the condensate and the peak of the susceptibility
always move to lower temperatures as $B$ is increased.
{The inflection point of the 
unrenormalized Polyakov loop, which is plotted
in Fig.~\ref{fig:poly_T_B}, shows a similar behavior.}
{It is interesting to notice, especially looking at the behavior 
of the chiral susceptibility peak, that at the same time
the strength of the crossover transition seems to increase.}

\begin{figure}[t!]
\centering
\includegraphics*[width=1\columnwidth]{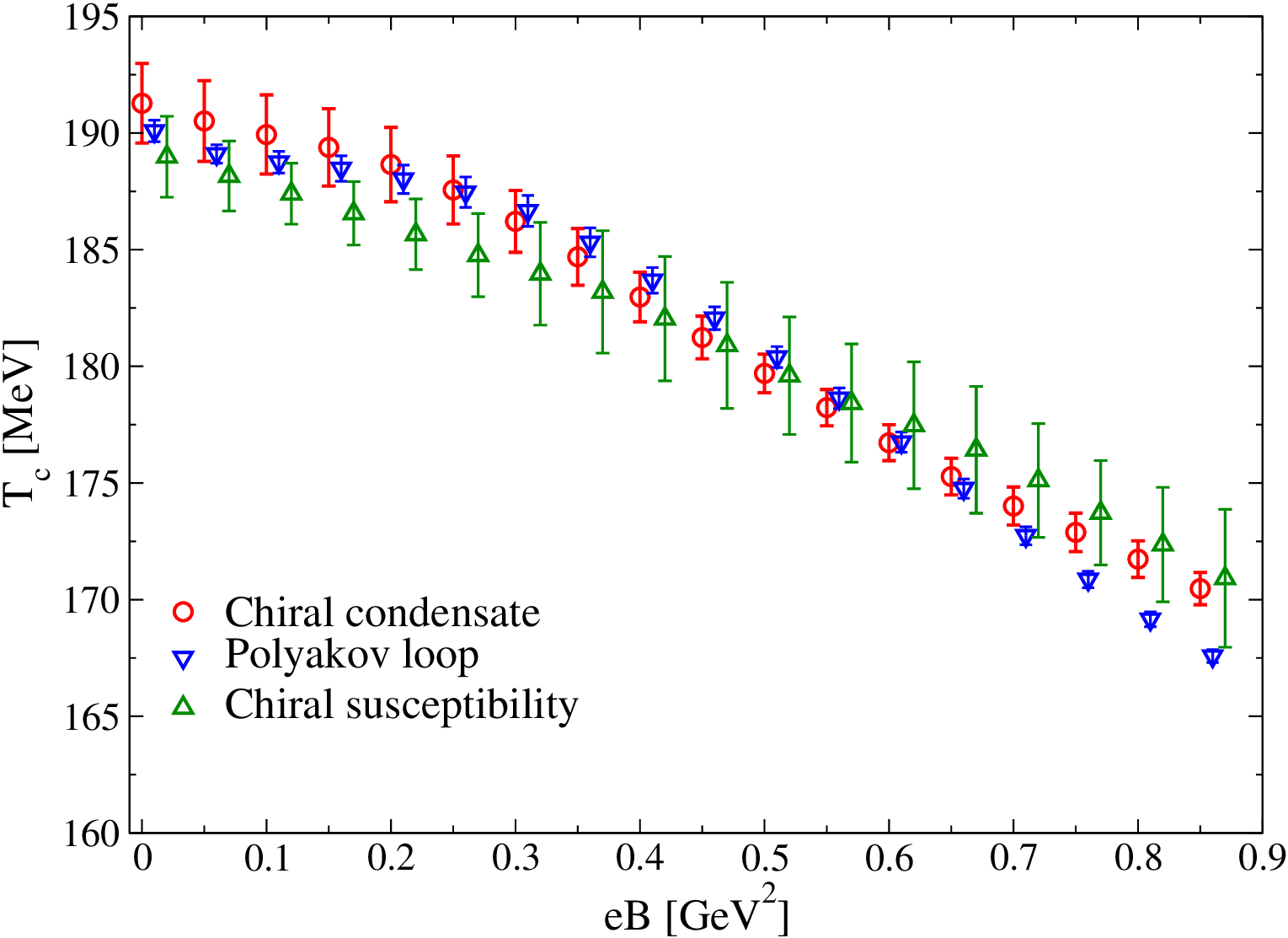}
\caption{Determinations of $T_c(B)$ obtained from different observables at $m_\pi=440\,\text{MeV}$. Data points have been shifted slightly to improve readability.
}
\label{fig:confronto}
\end{figure}

\begin{figure}[t!]
\centering
\includegraphics*[width=1\columnwidth]{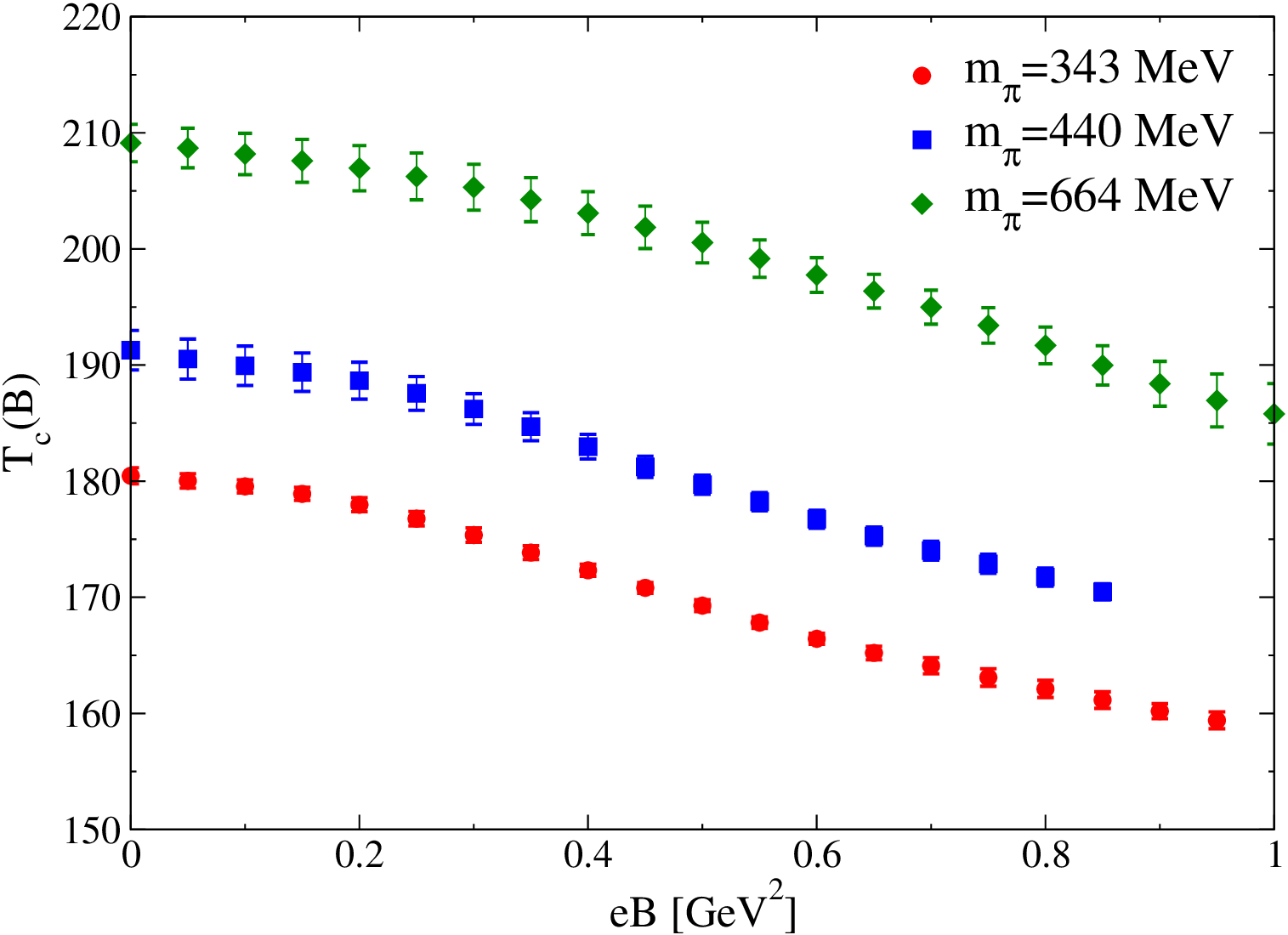}
\caption{Determinations of $T_c(B)$ obtained from the
{renormalized} condensate at the different pion masses.
}
\label{fig:tccond_B}
\end{figure}

For each value of $e B$, we have determined the location of 
$T_c$ using the same procedure (also regarding
the determination of the systematic error) already described {above},
i.e.~the inflection point for the {renormalized}
condensate and 
for the Polyakov loop, the maximum of the peak for the {chiral} 
susceptibility.
Results obtained for $m_\pi = 440$ and
for all different observables are reported in 
Fig.~\ref{fig:confronto}: the temperature decreases in a similar
way in all cases.
{In Fig.~\ref{fig:tccond_B} we report instead} the pseudo-critical
temperature obtained from the inflection point
of the {renormalized} condensate as a function of $eB$ for the different
values of the pion mass.

\begin{figure}[t!]
\centering
\includegraphics*[width=1\columnwidth]{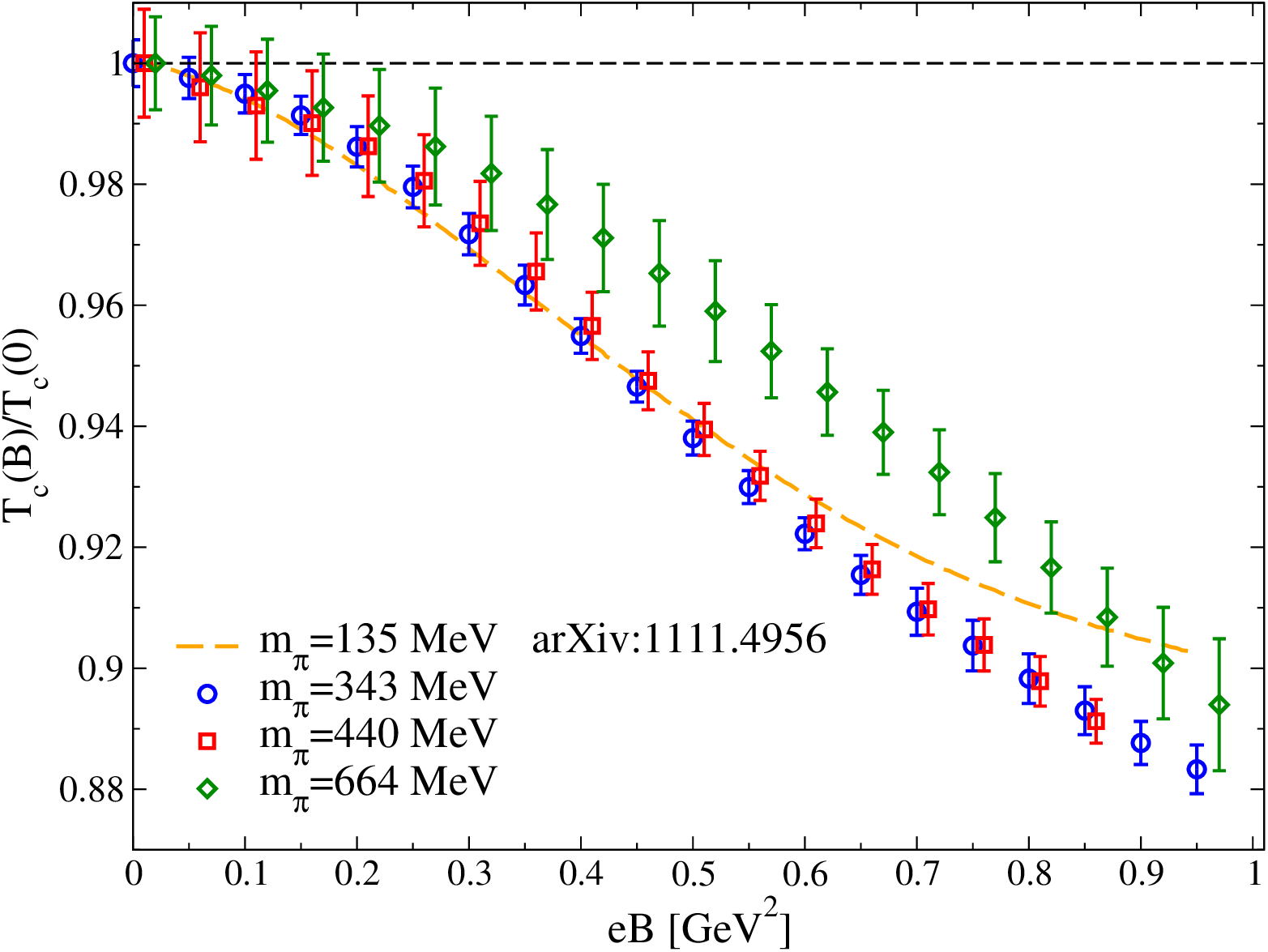}
\caption{$T_c(eB)/T_c(0)$ 
as a function of $eB$ for the different pion masses. Data points have been shifted slightly to improve readability. {Data at the physical point 
correspond to the results on $N_t = 6$ lattices reported in Ref.~\cite{reg0}.}
}
\label{fig:confronto_crescita}
\end{figure}

{It can be deduced that} the decrease of $T_c$ with $B$ {is} a general
phenomenon which, {from a qualitative point of view}, 
is independent of the quark mass {spectrum}.
However, some dependence {on $m_\pi$} 
must be present
at a quantitative level, because
the influence of the magnetic field on the gluon field
distributions takes place only through dynamical quarks,
so when their masses go to infinity the magnetic field decouples
and $T_c$ {must} 
become independent of $B$.
A tendency for a less pronounced influence as $m_\pi$ increases is 
qualitatively visible
in Fig.~\ref{fig:confronto_crescita}, where
we report $T_c(eB)/T_c(0)$ obtained
from the {renormalized} {quark} condensate for the different pion masses.

\begin{figure}[t!]
\centering
\includegraphics*[width=1\columnwidth]{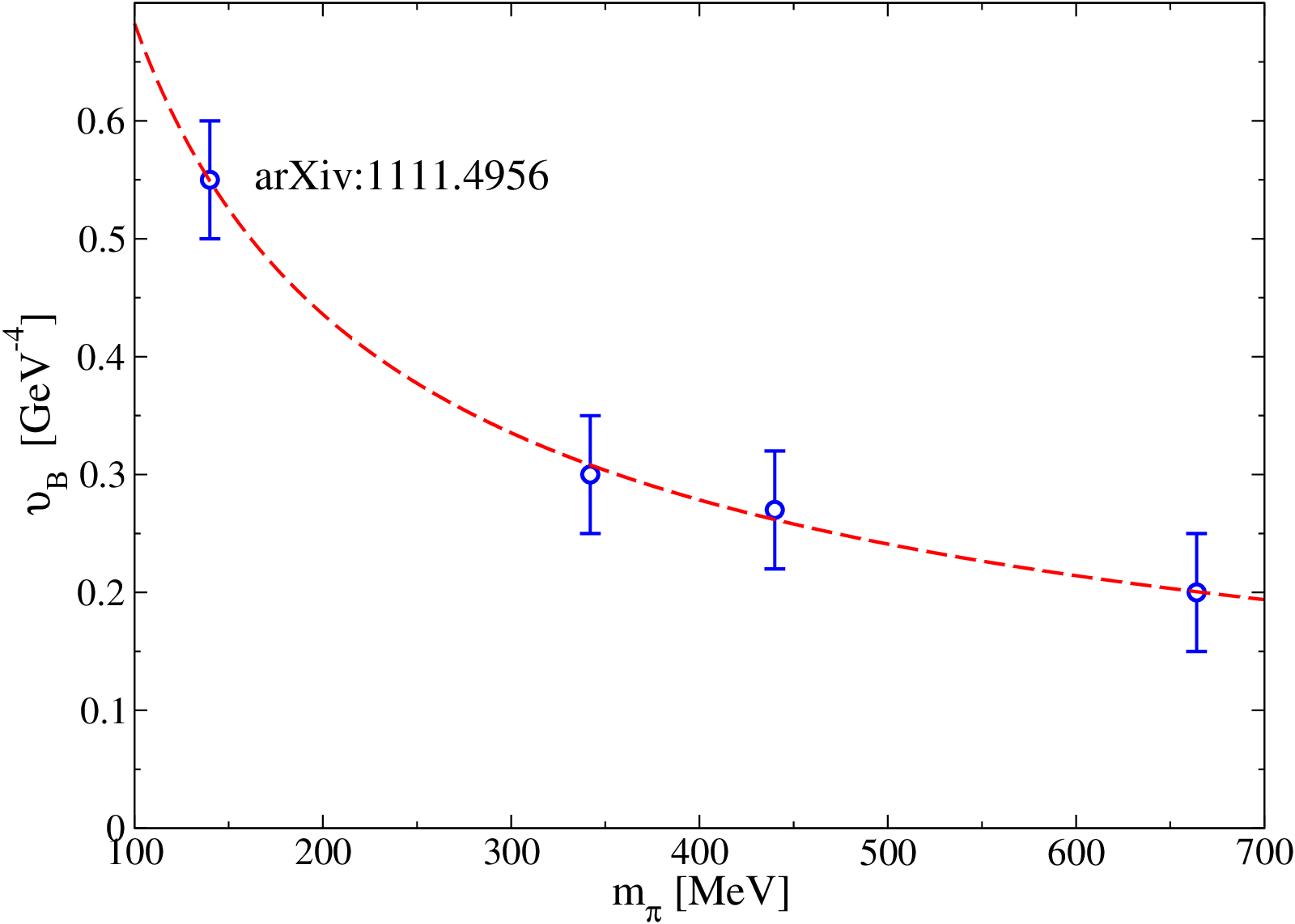}
\caption{Curvature coefficient $\upsilon_B$ defined in Eq.~(\ref{eq:curvature}) as a function of the pion mass.
}
\label{fig:curvature}
\end{figure}

{In} order to check this expectation on a quantitative
basis, we have tried to describe the small-$B$ dependence of $T_c$
in terms of a curvature coefficient
\beq
\frac{T_c(eB)}{T_c{0}} = 1 - v_B\, (e B)^2
\label{eq:curvature}
\eeq
obtaining the results reported in Table~\ref{tab:curvature} 
and in Fig.~\ref{fig:curvature}. The reported errors on $v_B$ take
into account the systematic uncertainty related to {the}
choice of the 
range over which {a best fit} 
according to Eq.~(\ref{eq:curvature}) is performed;
moreover, the value at the physical pion mass has been
obtained by fitting data for $T_c(e B)$ reported for $N_t = 6$
in Ref.~\cite{reg0}.
The dotted line in Fig.~\ref{fig:curvature} represents
the result of a best fit to a behavior $v_B = a/m_\pi^\alpha$
which yields $\alpha = 0.62(12)$ and serves just to prove that
data are compatible with a vanishing $v_B$ in the quenched limit,
however we stress that different behaviors,
obtained by fixing $\alpha = 1$ or choosing $v_B = a \exp(-m_\pi/b)$
work equally well and we are not able to fix the actual 
dependence of $v_B$, {due to} 
the present precision of our data.
{Moreover,} since we are considering data at fixed values of the cut-off
($N_t = 6$), the actual observed power law might be affected by the fact
that, with our lattice discretization, just one pion becomes
massless as the chiral limit is approached.

\begin{table}
	\caption{\label{tab:curvature} Curvature coefficient $\upsilon_B$ 
{defined in Eq.~(\ref{eq:curvature})} at the different pion masses.}
	\begin{ruledtabular}
		\begin{tabular}{ccc}
		$M_{\pi_0}\,[\text{MeV}]$& $\upsilon_B\,[\text{GeV}^{-4}]$ & $\Delta \upsilon_B\,[\text{GeV}^{-4}]$ \\ 
		\hline 
		135 & 0.55 & 0.05 \\ 
		342 & 0.30 & 0.05  \\
		440 & 0.27 & 0.05	\\		
		664 & 0.20 & 0.05 \\ 
		\end{tabular}
	\end{ruledtabular}
\end{table}

As a final remark, we {stress} that 
the results presented in this section show that
the discrepancy between the results reported 
in Refs.~\cite{demusa} and \cite{reg0} 
was just due 
to discretization effects, and not {the result of a different quark mass
spectrum.} This is also supported by Ref.~\cite{Tomiya:2017cey} where,
adopting the same unimproved discretization of Ref.~\cite{demusa},
$T_c$ continues to be an increasing function of $B$, {even 
for lighter-than-physical quark masses.}

\section{Inverse or direct magnetic catalysis around $T_c$}
\label{catalysis}

In QCD with physical quark masses,
the decreasing behavior of $T_c$ as a function of $B$ is 
associated with another interesting and 
unexpected phenomenon taking place around $T_c$: the enhancement
of chiral symmetry breaking, 
which is expected on general
grounds and actually observed at $T = 0$, is reversed around 
$T_c$, where instead $\pbp$ becomes a decreasing function of $B$.
This phenomenon has been given the name {\em inverse magnetic catalysis},
which has a clear correspondence with what happens.
However, with a less obvious correspondence, the same name 
has been usually assigned also to the decreasing behavior of $T_c$ itself.

Apart from merely lexical issues, the question is whether the observed
decrease of the condensate is actually the driving phenomenon
leading to the decrease of $T_c$, or if, on the contrary, 
some other phenomena force $T_c$ to decrease,
like those related to the influence of $B$ on the 
confining properties.
In this scenario, the observed behavior of $\pbp$ 
would be a secondary phenomenon: 
around the transition, when $B$ is switched on while
keeping the temperature $T$ fixed, 
the condensate decreases just 
because that temperature is moving into the deconfined and
chirally restored phase. 
However let us say that,
given our present ignorance about the relation existing 
between confinement and chiral symmetry
breaking, the question itself might be ill-posed.

\begin{figure}[t!]
\centering
\includegraphics*[width=1\columnwidth]{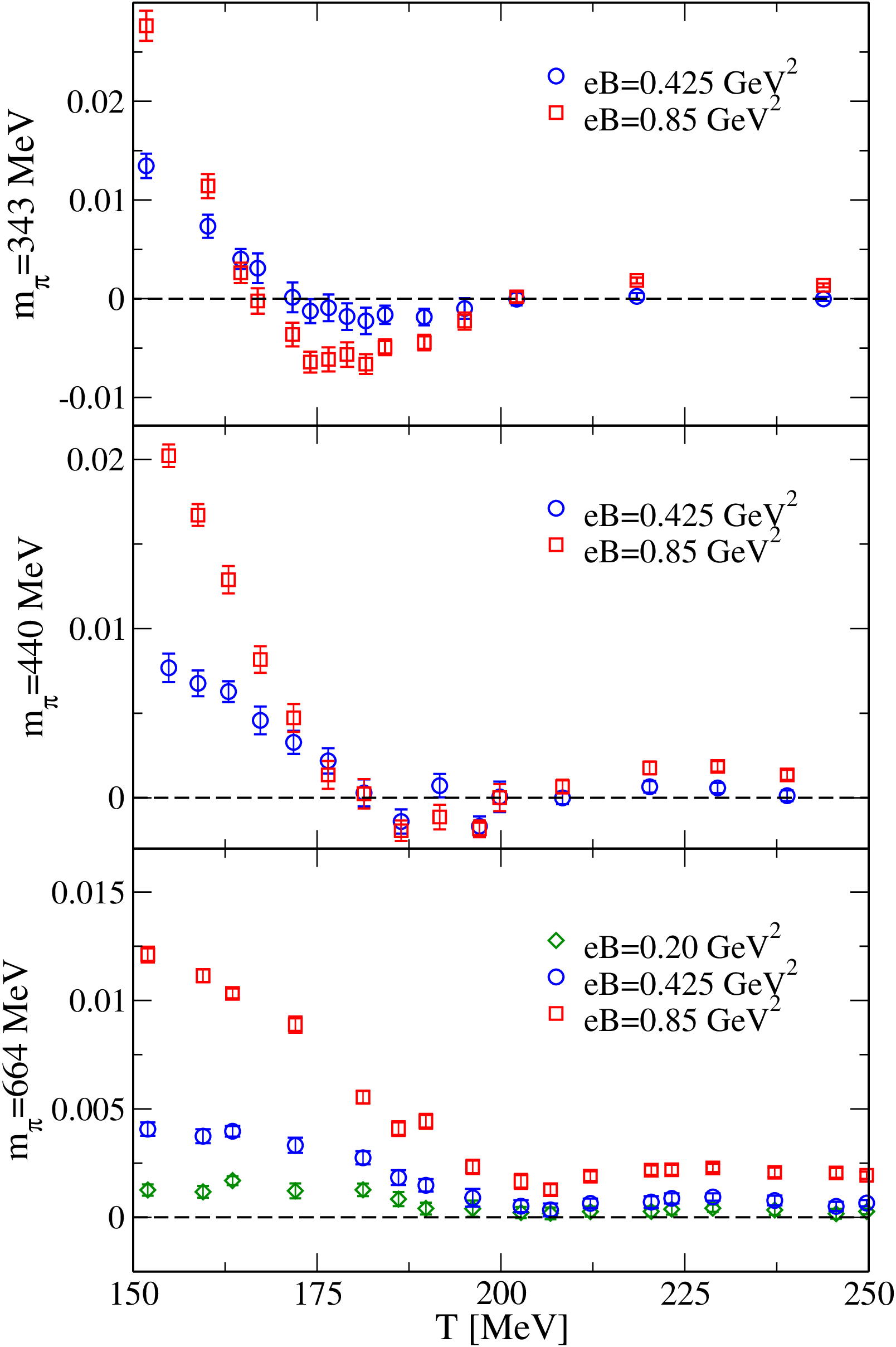}
\caption{Difference between the condensate at $B\neq0$ and $B=0$ as a function of $T$ at the different pion masses.
}
\label{fig:diff_cond}
\end{figure}

Nevertheless, from a practical point of view, we can investigate
if the relation between inverse magnetic catalysis 
and the decrease of $T_c$ is maintained 
also for larger-than-physical values of the pion mass.
To that purpose, in Fig.~\ref{fig:diff_cond} we report
the variation of the {quark} condensate, measured with respect to
its value at $B = 0$, as a function of the temperature
and for two different values of the magnetic field.  
Inverse magnetic catalysis is well visible, in a region
around $T_c$, for $m_\pi = 343$ MeV; however the effect becomes
barely visible for $m_\pi = 440$ MeV and completely disappears
for $m_\pi = 664$ MeV, where the condensate is always
an increasing function of $B$ at all temperatures. 
That gives evidence that one may have a decreasing behavior 
of $T_c$ even in absence of inverse magnetic catalysis,
challenging the strict connection, or even identfication, 
which has been usually assumed between the two phenomena.

\begin{figure}[t!]
\centering
\includegraphics*[width=1\columnwidth]{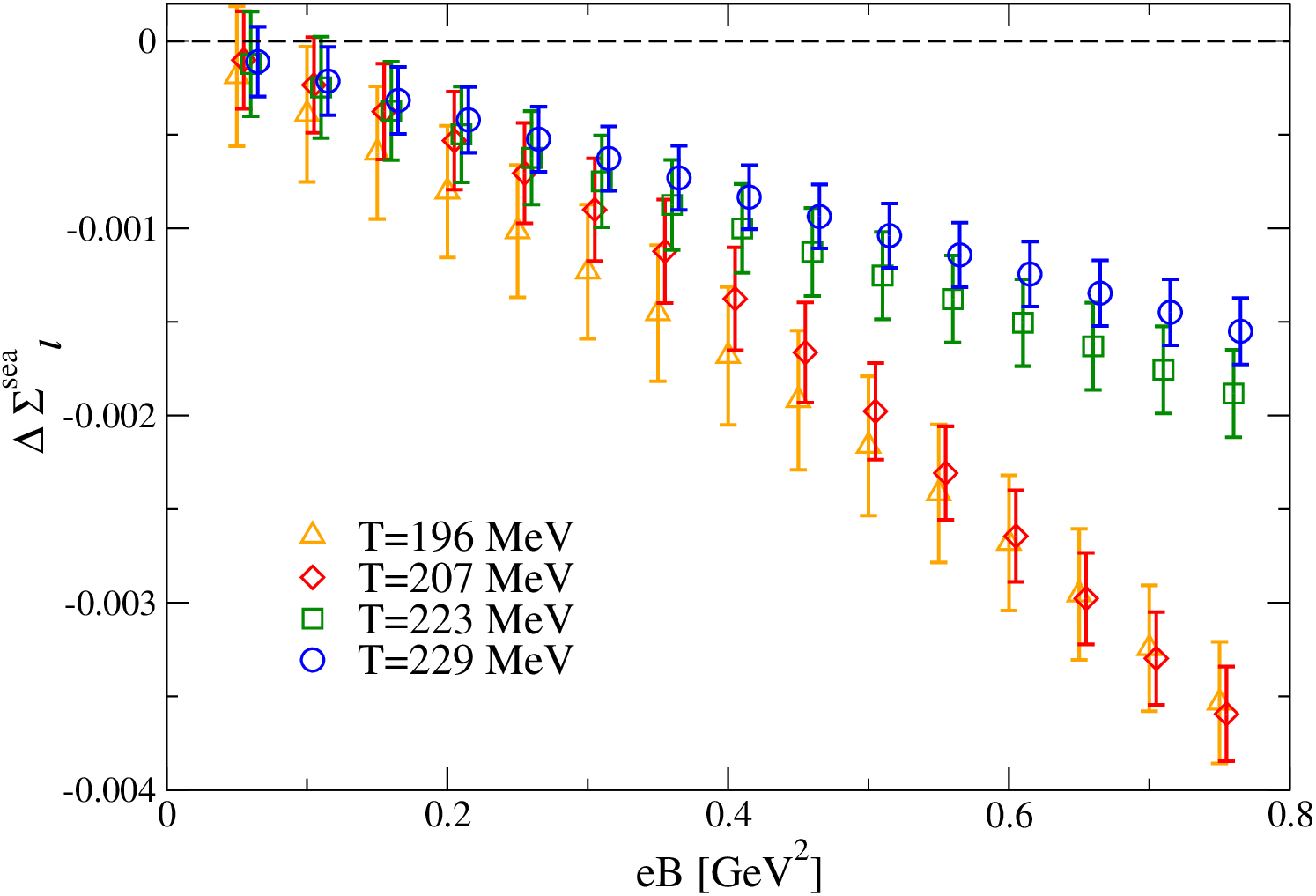}
\caption{Renormalized sea condensate as a function of $|e|B$ at $m_\pi=664\,\text{MeV}$ for various temperatures around the transition. Data points have been shifted slightly to improve readability.}
\label{fig:sea_up}
\end{figure}

{On the other hand}, when one considers the separation of the modification
of the {quark} condensate into a valence and sea 
contribution, {one still continues to 
see a behavior consistent with 
inverse magnetic catalysis in the sea contribution only and
for all values of the explored pion masses. This is visible in 
Fig.~\ref{fig:sea_up}, where we report the sea contribution
$\Delta\Sigma^{sea}_\ell(T,B)$
defined in Eq.~(\ref{eq:deltaseacond})}
for the largest pion mass and for a set of temperatures around the 
transition. 

{The fact that the sea contribution still continues
to be a decreasing function of $B$ around $T_c$
is of course related to how
the gauge field distribution gets
modified by $B$, so as to move towards (or deeper into)
the deconfined/chirally {restored} phase. As a consequence, the distribution
of eigenvalues of the $B = 0$ Dirac operator changes, and small
eigenvalues are suppressed.}

'

\section{Discussion and Conclusions}
\label{conclusions}

We have investigated the modification of the 
pseudo-critical temperature of QCD with $N_f = 2+1$ flavors
as a magnetic background field is switched on. The study 
has been done on lattices with a fixed temporal extent, $N_t = 6$,
changing the temperature along lines of constant physics corresponding to
three different values of the pseudo-Goldstone pion mass,
 $m_\pi = 343, 440$ and $664$ MeV.

We have found that the pseudo-critical temperature has always a 
decreasing behavior as a function of $eB$, even if the relative
variation, measured in terms of the curvature coefficient
$v_B$ defined in Eq.~(\ref{eq:curvature}), appears to be a decreasing
function of $m_\pi$ approaching zero in the quenched limit, as expected.
At the same time, we have observed that inverse magnetic catalysis 
disappears at all temperatures for the largest value of the pion mass,
even if it is maintained in the sea contribution only.
{Finally, we have observed that the magnetic 
field induces a strengthening of the crossover,
which is agreement with previous lattice 
studies}~\cite{demusa, reg0,Tomiya:2017cey} {as well
as with the prediction for a first order transition
at large enough magnetic field strengths}~\cite{Cohen:2013zja,Endrodi:2015oba}.
Present results are limited to a single value of the 
temporal extension, $N_t = 6$, therefore the exploration
of finer lattice spacings would be welcome in the future.

The persistence of the 
decreasing behavior of $T_c(B)$ 
observed even for large values of the pion mass,
where the chiral properties of the theory are not relevant,
{and possibly up to the quenched limit,
clarifies definitely the origin of the discrepancy 
between} Refs.~\cite{demusa} and \cite{reg0}
{and sheds some light on the origin of this phenomenon.
The fact that it is qualitatively independent 
of the quark mass spectrum is in agreement with
model computations which have found evidence of it
in a large-$N_c$ framework} (see, e.g., Refs.~\cite{icm0,dudal15}).
{The fact that it is not necessarily associated with a decrease of the 
quark condensate as a function of $B$ would
suggest to name the phenomenon as deconfinement 
catalysis~\cite{Bonati:2016kxj} rather than inverse magnetic
catalysis.}

{The paramagnetic behavior of the deconfined phase of strongly
interacting matter has been sometimes suggested
as a possible origin of this deconfinement catalysis.
Actually, such paramagnetic behavior has been observed even
when adopting the same unimproved lattice action leading 
to an increase of $T_c$}~\cite{demusa}, however the results
of Refs.~\cite{Bonati:2013lca,Bonati:2013vba} {show that
coarse and unimproved discretizations tend to make
the paramagnetic behavior weaker, especially around the 
pseudo-critical temperature.}

{The direct effects on the confining properties of the theory,
which are observed both a zero and at finite temperature, 
are other natural candidates to explain the deconfinement catalysis.
At finite temperatures below $T_c$ the magnetic field 
suppresses the string tension, while strong
enough magnetic fields could even lead to 
an anisotropic deconfinement at $T = 0$~\cite{Bonati:2016kxj}.}
{It would be interesting, in the future, to investigate
if, analogously to what happens for the behavior of $T_c$,
such effects on the confining properties persists also
for larger-than-physical quark masses and possibly towards the 
quenched limit. At the same time, in view of the observed 
strengthening of the transition induced by the magnetic field,
it would be interesting to investigate if the first order region
which is found around the quenched limit is enlarged 
by the presence of the magnetic background.}
\\

\acknowledgments
Numerical simulations have been performed on the MARCONI
machine at CINECA (Project Iscra-B/IsB16\_TCBQCD), 
on the COKA cluster at INFN-Ferrara
and at the Scientific Computing
Center at INFN-Pisa. 
FN acknowledges financial support from the INFN HPC\_HTC project.

\end{document}